\documentclass{elsart}
\usepackage{epsfig}

\def\lsim{\mathrel{\rlap{
\lower4pt\hbox{\hskip-3pt$\sim$}}
    \raise1pt\hbox{$<$}}}     
\def\gsim{\mathrel{\rlap{
\lower4pt\hbox{\hskip1pt$\sim$}}
    \raise1pt\hbox{$>$}}}     

\journal{Physics Letters B}

\begin{document}

\begin{frontmatter}

\title{Thermodynamics of $\Delta$ resonances}

\author{W.~Weinhold\thanksref{email}},
\author{B.~Friman},
\author{W.~N{\"o}renberg}
\address{ Gesellschaft f\"ur Schwerionenforschung (GSI),\\
Planckstr. 1, D-64291 Darmstadt, Germany \\
\and\\
Institut f{\"u}r Kernphysik, Technische Universit\"at Darmstadt,\\
Schlo{\ss}gartenstr. 9, D-64289 Darmstadt, Germany}
\thanks[email]{e-mail: w.weinhold@gsi.de}

\begin{abstract}
The thermodynamic potential of a system of pions and nucleons is
computed including the $\pi$N interactions in the $P_{33}$ channel.
A consistent treatment of the width of the resonance in this
channel, the $\Delta(1232)$ resonance, is explored in detail. In
the low-density limit we recover the leading term of the virial
expansion for the thermodynamic potential.  An instructive
diagrammatic interpretation of the contributions to the total
baryon number is presented. Furthermore, we examine within a
fireball model the consequences for the pion spectra in heavy-ion
collisions at intermediate energies, including the effect of
collective flow. A consistent treatment of the $\Delta$ width leads
to a substantial enhancement of the pion yield at low momenta.
\end{abstract}
\end{frontmatter}

\section{Introduction}

In heavy-ion collisions a zone of hot and dense hadronic matter is
created. At beam energies on the order of 1~GeV per nucleon or
higher, hadron resonances play an important role in the dynamics of
such collisions. Some of these resonances have large widths,
comparable to or larger than the typical temperature of the
interaction zone. Also stable hadrons generally acquire a width in
dense matter, due to the interactions with the medium.
Consequently, a consistent theory of the thermodynamics and
kinetics of many-body systems with short-lived particles is
necessarily a basic element in the description of such collisions.
At present a satisfactory treatment of this problem is still
lacking.

In this letter we explore the consequences of the width of the
$\Delta(1232)$ resonance on the thermodynamics of hot and dense
hadronic matter. In heavy-ion collisions at beam energies around 1~GeV
per nucleon, the nucleon and the pion are together with the
$\Delta(1232)$ the most abundant hadrons. By treating the
$\Delta(1232)$ as an independent field, one effectively accounts for
the pion-nucleon interactions in the resonant $P_{33}$
channel. As a first approximation we assume that the interaction zone
consists only of these particles. The extension of our approach to
other channels is straightforward.

The thermodynamics of such systems has been studied by many
authors. In calculations based on the Green-function or related
techniques the width of the $\Delta$ was, so far, either neglected
or dealt with in an {\em ad hoc} manner
\cite{woGamma,wGamma,GorTsaiYang95,Dani95}. On the other hand it is
well known how to take interactions in general and resonances in
particular into account within the S-matrix formulation of
statistical mechanics \cite{DMB,exdelta}. In section 2 we
illustrate the relation between the two techniques within a 
model.

In order to assess the consequences of the $\Delta$ width for the pion
spectrum in heavy-ion collisions we consider a simple fireball model
(section 3). In this model it is assumed that the interaction region,
created in such collisions, is in local thermal and chemical
equilibrium at freeze-out, i.e., when the particles decouple.
There are two sources for the observed pions:
the free pions with a thermal distribution present at freeze-out and the
pions emanating from the decay of thermally excited
$\Delta$-resonances. We show that popular prescriptions for evaluating
the resonance contribution are incorrect.

\section{Thermodynamic potential and baryon density}

Our starting point for computing the thermodynamic potential
$\Omega$ of a $\pi$N$\Delta$-gas as a function of temperature $T$,
baryon chemical potential $\mu_B$ and volume $V$ is the Green-function
approach of Baym and Kadanoff \cite{Baym}.
The temperature Green functions are denoted by ${\mathcal{G}}_N$,
${\mathcal{G}}_\Delta$ for the nucleon and $\Delta$ resonance
respectively, and by ${\mathcal{D}}_\pi$ for the pion.  The free
Green functions are defined by
\begin{eqnarray}
{\mathcal{G}}^0_B  \left( \pol{p},E_m \right)&  = &\left[
                \left( i E_m - \mu_B \right)
         - \epsilon_B \left( \pol{p} \right) \right]^{-1},
          \qquad
          \epsilon_B \left( \pol{p} \right) =
          m^0_B+ \pol{p}\,{}^2 / 2 m^0_B \label{baryon}\\
{\mathcal{D}}^0_\pi  \left( \pol{q} , \omega_m \right) & = &
                  \left[ \left( i \omega_m \right)^2
                 - \left( \omega_\pi \left( \pol{q} \right) \right)^2
                \right]^{-1} ,
          \qquad
          \omega_\pi \left( \pol{q} \right) =
          \sqrt{ \pol{q}\,{}^2 + {m^0_\pi}^2 }\label{pion}\,,
\end{eqnarray}
where $B=N,\Delta$ while $E_m$ and $\omega_m$ denote the Matsubara
frequencies for fermions and bosons, respectively. We consider an
interaction vertex of the $\pi N \Delta$ type, which connects a
pion and a nucleon with a $\Delta$ resonance. The following
discussion depends only on the topology of the diagrams considered
and not on a particular choice of the coupling. Hence, we specify
the form of the interaction only at the end of this section, where
numerical results are presented.

The exact thermodynamic potential $\Omega$ of an interacting system
of pions, nucleons and $\Delta$ resonances can be expressed in
terms of the full Green functions~\cite{Baym,formPhi,exPhi}
\begin{eqnarray}
\Omega & = & \frac{1}{2} \; \mbox{Tr} \; \mbox{ln}\left[ -
{\mathcal{D}}_\pi^{-1} \right]
+ \frac{1}{2} \; \mbox{Tr} \; \Pi_\pi {\mathcal{D}}_\pi
- \mbox{Tr} \; \mbox{ln}\left[-{\mathcal{G}}_N^{-1} \right]
 - \mbox{Tr} \; \Sigma_N {\mathcal{G}}_N
\nonumber\\
\label{start}
& & - \mbox{Tr} \; \mbox{ln}\left[-{\mathcal{G}}_\Delta^{-1} \right]
    - \mbox{Tr} \; \Sigma_\Delta {\mathcal{G}}_\Delta
    + \Phi \,,
\end{eqnarray}
where $\Sigma_N$, $\Sigma_\Delta$ and $\Pi_\pi$ are the self
energies of the nucleon, the $\Delta$ resonance and the pion. The
traces correspond to sums over spin, isospin, and Matsubara
frequencies as well as integrations over phase space. The functional
$\Phi$ is the sum of all two-line irreducible skeleton diagrams
where the lines represent full Green functions. The self energy of
a particle is obtained by a functional variation of $\Phi$ with
respect to the Green function: $\Sigma_B =\delta \Phi /\delta
{\mathcal{G}}_B$ and $\Pi_\pi = -2\;\delta \Phi / \delta
{\mathcal{D}}_\pi$. Diagrammatically, the functional variation  is
accomplished by opening the corresponding line~\cite{formPhi}.

\begin{figure}[tb]
\unitlength1mm
\begin{center}\begin{picture}(120,27)
\put(-2,21){(a)}
\put(0,3){\makebox{
 \epsfig{file=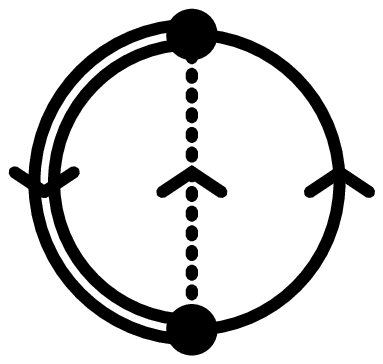,height=20mm,width=20mm}}}
\put(38,21){(b)}
\put(40,3){\epsfig{file=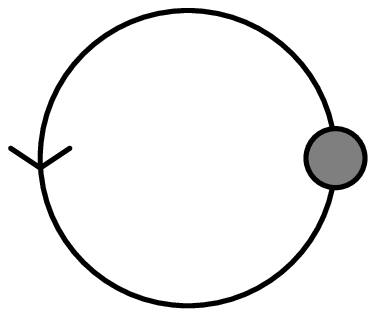,
        height=20mm, width=20mm}}
\put(63,12){+}
\put(70,3) {\epsfig{file=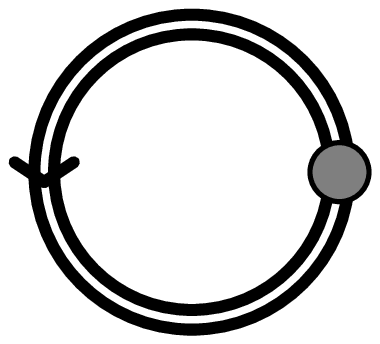,
        height=20mm, width=20mm}}
\put(94,12){+}
\put(100,3) {\epsfig{file=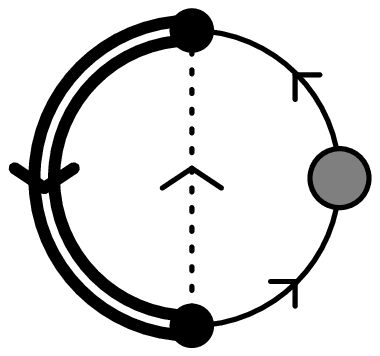,
        height=20mm, width=20mm}}
\end{picture}\end{center}
\caption[]{Diagrammatic representation of (a) the functional $\Phi$
and (b) contributions to the baryon density. A double line denotes a
$\Delta$, a solid line a nucleon and a dashed line a pion Green
function.  Thick lines correspond to dressed, thin lines to free Green
functions. The gray circle indicates the baryon number operator. The
first diagram in (b) yields the nucleon density $n_N$, the second one
the density of $\Delta$ resonances in the system $n_\Delta$, while the
third one, $\delta n_N$, corresponds to the $\pi$N component of a
physical $\Delta$ (see eqs.~(8) and (9)). \\}
\label{functional}
\end{figure}

As shown by Baym and Kadanoff \cite{Baym}, the choice of a subset
of diagrams for $\Phi$ defines a self-consistent approximation,
where every Green function is dressed by its self energy obtained
by the variation of $\Phi$. The advantage of such so-called
conserving or $\Phi$-derivable approximations is that they satisfy
conservation laws and guarantee thermodynamic consistency. This
approach has in the past been applied to a wide variety of
many-body problems \cite{exPhi}. For the $\pi$N$\Delta$ system the
simplest approximation of this type is obtained by choosing $\Phi$
equal to the two-loop diagram shown in fig.~\ref{functional}(a).
The corresponding self energy of the $\Delta$ resonance is obtained
by opening the $\Delta$ line.

A fully self-consistent treatment of this problem would be quite
involved, and outside the scope of this exploratory calculation. We
therefore relax the self-consistency and consider a low-density
expansion\footnote{Strictly speaking we loose the nice properties of
the conserving approximations in this step. However, at least for the
problem at hand, the leading term in the low-density expansion is
close to the self-consistent solution at moderate
densities~\cite{next} $(\rho\lsim 2\rho_0$). Self consistency is
expected to be important, when there is a small scale in the problem,
see e.g.  ref.~\cite{matthias}.} of eq.~(3). In the pion and nucleon
Green functions we retain only the pole terms, i.e., we use
eqs. (1-2), albeit with physical masses. On the other hand, for the
$\Delta$, which has a large width, the vacuum self energy
$\Sigma_\Delta$ is treated explicitly. One then finds, to leading
order in the baryon density
\begin{eqnarray}
\label{now}
\Omega   =
    \frac{1}{2} \; \mbox{Tr} \; \mbox{ln}\Bigl[ -
\left( {\mathcal{D}}_\pi^V \right)^{-1} \Bigr]
- \mbox{Tr} \; \mbox{ln}\Bigl[- \left( {\mathcal{G}}_N^V
\right)^{-1}
         \Bigr]
- \mbox{Tr} \; \mbox{ln}\Bigl[-\left(
               {\mathcal{G}}_\Delta^V \right)^{-1} \Bigr] \,,
\end{eqnarray}
where ${\mathcal{D}}_\pi^V = {\mathcal{D}}_\pi^0$ and
${\mathcal{G}}_N^V = {\mathcal{G}}_N^0$ with $m_\pi^0
= m_\pi$ and $m_N^0 = m_N$, respectively and $\left({
\mathcal{G}}_\Delta^V\right)^{-1} = \left({
\mathcal{G}}_\Delta^0 \right)^{-1} - \Sigma_\Delta$. Since the
thermodynamic potential in eq.~(\ref{now}) is a sum of independent
terms involving the vacuum Green functions, it may superficially
appear that the problem has now been reduced to a trivial one with
non-interacting pions, nucleons and $\Delta$ resonances. However,
as we show below, this approximation includes non-trivial
interaction effects. In fact, the $\Delta$-resonance term accounts
for the $\pi$N interactions in the $P_{33}$ channel.

The traces in eq.~(\ref{now}) can be evaluated using standard
techniques (cf. \cite{FetWal71}). For a spin and isospin symmetric
system one finds

\begin{eqnarray}
\Omega \left( T, \mu_B, V \right) & = &
      3 \; T V\; \int \frac{{\d} ^3 q}{\left( 2 \pi \right)^3} \;
     \mbox{ln} \left[ 1 - {\e}^{-\beta
      \omega_\pi \left( \pol{q} \right)} \right] \nonumber \\
&&  -4 \; T V\;
        \int \frac{{\d} ^3 p}{\left( 2 \pi \right)^3} \;
     \mbox{ln}
        \left[ 1 + {\e}^{-\beta \left( \epsilon_N \left( \pol{p} \right)
                - \mu_B \right)} \right] \nonumber \\
&& \label{thermopot1}
 - 16  \; T V\; \int \frac{{\d} ^3 p}{\left( 2 \pi \right)^3}
    \; \int\limits^{\infty}_{-\infty}
      \frac{{\d} E}{2 \pi} \; \mathcal{B} \left( \pol{p}, E \right) \;
      \mbox{ln} \left[1 + {\e}^{ - \beta \left( E - \mu_B \right)}
      \right] ,
\end{eqnarray}
where the first term is the thermodynamic potential of a free pion
gas and the second one is that of a free nucleon gas.  The last
term in eq.~(\ref{thermopot1}) accounts for the interaction
contribution to the thermodynamic potential. It is similar to the
thermodynamic potential of a free Fermi gas, but involves an
integration over the energy with a weight function
\begin{eqnarray}
\mathcal{B} \left( \pol{p},E \right) & = &  - 2 \; \mbox{Im}
{\Bigg[} \left( 1 -
\frac{\partial \Sigma_\Delta^R\left( \pol{p},E \right)}{\partial E}
\right) \;
G_\Delta^R \left( \pol{p},E \right) {\Bigg]} \nonumber \\
\label{firstform}
& = &
\mathcal{A} \left( \pol{p},E \right) + 2 \; \mbox{Im}
{\Bigg[}
\frac{\partial \Sigma_\Delta^R\left( \pol{p},E \right)}{\partial E} \;
G_\Delta^R \left( \pol{p},E \right) {\Bigg]} \,.
\end{eqnarray}
Here $\Sigma_\Delta^R$ is the retarded self energy and
$\left(G_\Delta^R\right)^{-1}=\left( G_\Delta^0\right)^{-1}-
\Sigma_\Delta^R$ the retarded vacuum Green function of the $\Delta$
resonance, while $\mathcal{A}$ denotes its spectral function.  The functions
$\mathcal{A}$ and $\mathcal{B}$ fulfill the same sum rule, i.e., the
integral over all energies equals $2\pi$.

For an interpretation of the weight function $\mathcal{B}$ in terms of
diagrams it is instructive to consider the baryon density
\begin{equation}
n_B \left( T, \mu_B \right)  = - \;
\frac{1}{V} \;  \frac{\partial \; \Omega \left( T, \mu_B, V \right)}
{\partial \mu_B} {\Bigg |}_{T, V} \,,
\end{equation}
which can be split into the nucleon contribution
\begin{equation}
n_N \left(T, \mu_B \right)   =
   4 \; \int \frac{ {\d} ^3 p } {\left( 2 \pi
\right)^3} \; \frac{1}{{\e}^{\beta \left(
        \epsilon_N \left( \pol{p} \right)  - \mu_B
\right) } +1} \qquad
\end{equation}
and the contribution due to the pion-nucleon interactions
\begin{equation}
\label{states}
\tilde{n}_\Delta \left( T, \mu_B \right)  =
 16 \; \int \frac{
{\d} ^3 p } {\left( 2 \pi \right)^3} \; \int\limits_{-\infty}^{\infty}
\frac{ {\d}  E }{2 \pi} \; \mathcal{B} \left( \pol{p}, E \right) \;
\frac{1}{{\e}^{\beta \left( E - \mu_B \right) } + 1} \,.
\end{equation}
Furthermore, as implied by the form (\ref{firstform}) for
$\mathcal{B}$, the interaction contribution (\ref{states}) to $n_B$
consists of two parts. Both of these have a straightforward
interpretation: The first part, with the spectral function
$\mathcal{A}$, is the contribution $n_\Delta$ of the $\Delta$
resonance to the baryon density.  The second part, which is
proportional to $\partial\Sigma_\Delta/\partial E$, is the
contribution of the nucleons appearing in conjunction with a pion in
the $\Delta$ self energy. We denote this $\pi$N pair contribution by
$\delta n_N$.

We stress that there is no double counting here. The state vector
of an energy eigenstate in the interacting system has amplitudes
corresponding to a bare $\Delta$ and to the non-interacting $\pi$N
states. The spectral function projects the interacting states onto
the bare $\Delta$ state, while the second term in
eq.~(\ref{firstform}) represents the projection onto the $\pi$N
pair states\footnote{\samepage{A detailed analysis~\cite{Weinhold}
reveals that the second term not only accounts for the interacting
$\pi$N pairs, but also {\em subtracts out} the contribution of the
uncorrelated ones. This is necessary to avoid double counting,
since the contribution of uncorrelated nucleons and pions is taken
into account by the first two terms in eq.~(\ref{thermopot1}).}}.
Since the contribution of an energy eigenstate to the partition sum
depends only on its energy and not on the admixture of a particular
unperturbed state, both contributions in eq.~(\ref{firstform}) must
be included in the thermodynamic potential, eq.~(\ref{thermopot1}).

One advantage of the formulation in terms of Green functions is
that different contributions are easily interpreted in terms of
diagrams. The three contributions to the baryon density correspond
to the three diagrams shown in fig.~\ref{functional}(b). So far the
last contribution, $\delta n_N$, has been neglected in the
literature on calculations of the thermodynamics of $\Delta$
resonances \cite{wGamma,GorTsaiYang95,Dani95}. In such
calculations an {\em ad hoc} prescription was employed, where the
spectral function $\mathcal{A}$ was used in lieu of $\mathcal{B}$.

In order to make contact with earlier work \cite{DMB,BethUl}, we show
that $\mathcal{B}$ can be expressed in terms of the $\pi$N-scattering
phase shift in the $P_{33}$ channel.  First, we note that the direct
$\Delta$-Born graph is the only $\pi$N-scattering process that is
generated by the functional fig.~\ref{functional}(a).  Hence, the
$\pi$N-scattering amplitude is proportional to the vacuum $\Delta$
Green function $G_\Delta$ and the phase shift is given by\footnote{Due
to Galilean invariance the vacuum Green function and self energy
depend only on the energy in the $\pi N$ center-of-momentum frame
$E_{cm}$, i.e. $\Sigma_\Delta (\pol{p},E) = \Sigma_\Delta (E_{cm})$.}
\begin{equation}
\label{shifts}
\delta_{33} \left( E_{cm} \right) =
\arctan  \left[ \frac{ \mbox{Im}
     \Sigma_\Delta^R \left( E_{cm} \right)} {E_{cm} - m_\Delta^0 -
     \mbox{Re}\Sigma_\Delta^R \left( E_{cm} \right)} \right] \,,
\end{equation}
which implies that (cf. eq.~(\ref{firstform}))
\begin{equation}
\label{Bwithdelta}
\mathcal{B}\left( E_{cm} \right) =
2 \; \frac{\partial}{\partial E_{cm}}
\delta_{33} \left( E_{cm} \right) \,.
\end{equation}
This form of $\mathcal{B}$ coincides with that obtained in the
$S$-matrix formulation of statistical mechanics by Dashen, Ma and
Bernstein for narrow resonances \cite{DMB}. In the case we consider,
namely a broad resonance, there are corrections to this expression due
to interactions involving more than two particles. However, in the
low-density limit two-body scattering processes dominate
\cite{dover}. Thus, when the $\Delta$ self energy is properly
included, eq.~(\ref{now}) accounts for the $\pi$N interactions in the
$P_{33}$ channel to leading order.  When the logarithm in the last
term of eq.~(\ref{thermopot1}) is expanded in powers of
$\exp(-\beta(E-\mu_B))$, retaining only the lowest term, one recovers
the well-known expression for the second virial coefficient of Beth
and Uhlenbeck \cite{BethUl}.

The expression (\ref{Bwithdelta}) for the weight function
$\mathcal{B}$ allows an appealing interpretation of
eqs.~(\ref{thermopot1}) and (\ref{states}). The energy derivative of
the phase shift at the energy $E_{cm}$ can be identified with the time
delay of a scattered wave due to the presence of the scatterer
\cite{delay}. Hence, the contribution of a resonance to the
thermodynamic quantities is proportional to the time spent in the
resonant state \cite{DaniPratt96}.

\begin{figure}[t]
\unitlength1mm
\begin{center}\begin{picture}(120,135)
\put(0,-5){\epsfig{file=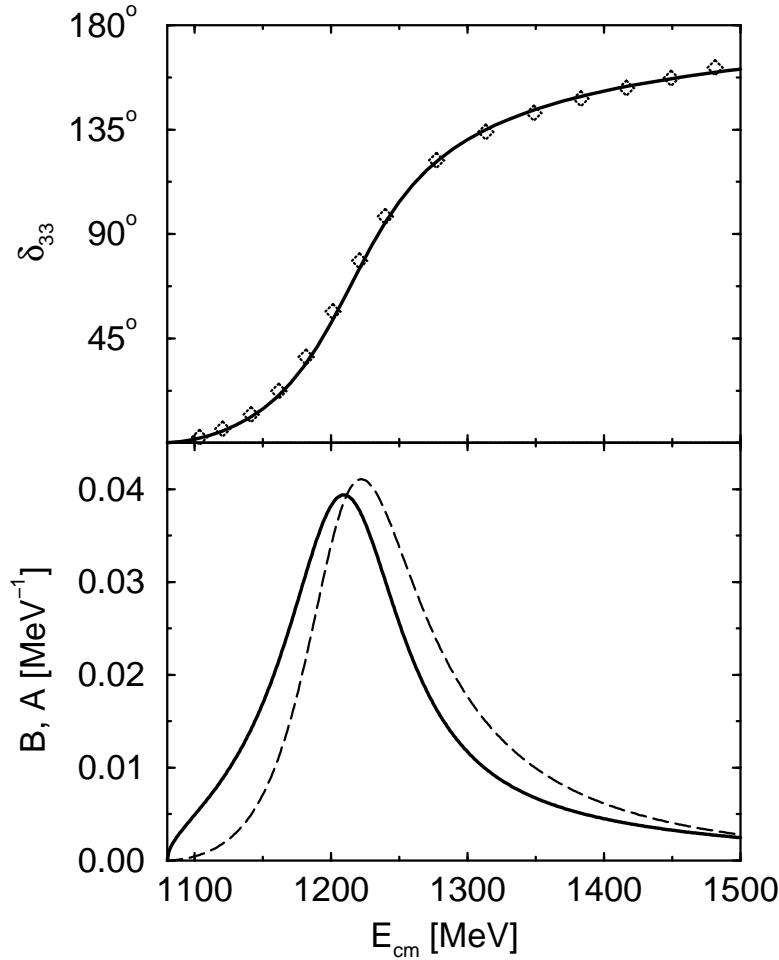,height=15cm}}
\end{picture}
\end{center}
\caption[...]{Upper part: Fit of $\delta_{33}$ from eq.~(\ref{shifts})
to the empirical $P_{33}$ phase shifts \cite{Arndt}; Lower part:
Weight function $\mathcal{B}$ (solid line), see eqs.~(\ref{firstform},
\ref{Bwithdelta}), compared to the spectral function $\mathcal{A}$
(dashed line).}
\label{phAB}
\end{figure}

The following numerical results were obtained using the $\pi
N\Delta$ interaction
$\left(f_{\pi\!N\!\Delta}/{m_\pi}\right) \left( \pol{q} \cdot \pol{S}
\right) T_a\;$
(cf. \cite{EricWeise}) with a form factor of the monopole type
$F(\pol{q}^{\,\, 2})=\Lambda^2/(\pol{q}^{\,\, 2}+\Lambda^2)$. The
parameters are chosen to reproduce the phase shift for $\pi$N
scattering in the $P_{33}$ channel \cite{Arndt} up to a
center-of-mass energy of 1.5 GeV (upper part of fig.~\ref{phAB}).
We find $f_{\pi\!N\!\Delta} = 3.28$, $\Lambda
= 290$ MeV and $m_\Delta^0 = 1316$ MeV.

In the lower part of fig.~\ref{phAB} we compare the weight function
$\mathcal{B}$ with the spectral function $\mathcal{A}$. This reveals a
qualitative difference close to threshold and a strong shift of
strength to lower masses, due to the second term of $\mathcal{B}$ in
eq.~(\ref{firstform}). The weight function $\mathcal{B}$ is uniquely
determined by the phase shifts, eq.~(\ref{Bwithdelta}), and hence a
model-independent quantity. On the other hand, the spectral function
and consequently the separation of $\tilde{n}_\Delta$,
eq.~(\ref{states}), into contributions from $\Delta$ resonances and
$\pi$N pairs are model dependent.

\section{Pion spectra}

In order to compare our calculation with experimental data we assume
that at freeze-out the fireball created in a heavy-ion collision is in
local thermal and chemical equilibrium.  Given the population of
states $\tilde{n}_\Delta$, which decay into a nucleon and a pion after
freeze-out, it is straightforward to compute the resulting pion
spectrum, cf. \cite{GorTsaiYang95,Hagedorn}.  The pion momentum
distribution in the $\Delta$ rest frame is folded with the resonance
velocity distribution due to thermal motion and collective flow.  The
collective radial flow \cite{Dani95} is taken into account by assuming
a flow profile, which depends quadratically on the radial
coordinate\footnote{This choice is not crucial, because we find only a
weak dependence of the final pion spectrum on the assumed analytic
form of the flow profile, for a given value of the mean flow
velocity.}.

The final $\pi^0$ spectrum is then a function of the temperature
$T$, the chemical potentials for baryon number $\mu_B$ and electric
charge $\mu_Q$ as well as the mean flow velocity $\bar{v}$. The
parameters at freeze-out are determined by fitting the $\pi^0$
spectrum \cite{Schwalb94} of the reaction Au+Au at E$_{lab}$/A = 1
GeV with the constraints that the ratio of charge to baryon number
equals ${N_Q}/{N_B} = 79/197$ and that the ratio of $\pi^+$ mesons
to the number of participating baryons is given by \cite{Senger96}
$N_{\pi^{+}_{obs}}/{N_B} \simeq 0.025$.
\begin{figure}
\unitlength1mm
\begin{center}\begin{picture}(120,133)
\put(0,-5){\epsfig{file=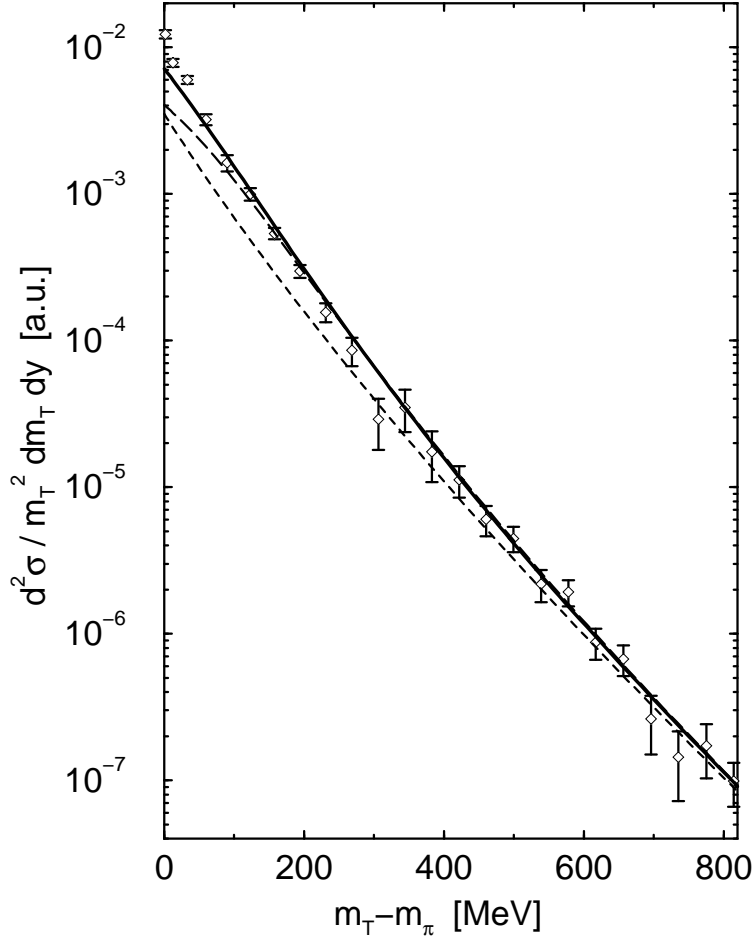,height=15cm}}
\end{picture}
\end{center}
\caption[...]{ The $\pi^0$ spectrum of Au+Au at E$_{lab}$/A = 1 GeV of
ref. \cite{Schwalb94}. The dashed line shows the contribution of
thermal pions, while the long-dashed line corresponds to thermal pions
plus those which stem from $\Delta$ resonances, decaying after
freeze-out. The latter contribution is determined by the spectral
function $\mathcal{A}$, fig.~\ref{phAB}. Finally, the full line
includes in addition the contribution of the $\pi$N pairs.\\}
\label{TAPS}
\end{figure}

In fig.~\ref{TAPS} we show the different contributions to the spectrum
as a function of the transverse pion mass $m_\perp = \sqrt{p_\perp^2 +
m_\pi^2}$ in a representation, where a Boltzmann distribution would
appear as a straight line with slope $-T^{-1}$.  At large $m_\perp$,
the spectrum is dominated by thermal pions, while the $\pi$N pairs
give an important contribution to the pion spectrum at small
transverse masses. Since the weigth function $\mathcal{B}$, which
determines the sum of the $\Delta$ and $\pi$N-pair contributions, is
model independent, the full line is not affected by assumptions
concerning the structure of the $\Delta$ resonance in vacuum, provided
the phase shifts are reproduced. At freeze-out we find a temperature
of $T= 57 \; \mbox{MeV}$, a baryon chemical potential of $\mu_B = 830$
MeV, a charge chemical potential of $\mu_Q = -18$ MeV and a mean flow
velocity of $\bar{v} = 0.36$. The corresponding freeze-out density is
$n_B = 0.4\;\rho_0$, where $\rho_0 = 0.16$ fm$^{-3}$ denotes normal
nuclear matter density.

As mentioned above the separation of $\tilde{n}_\Delta$ into a bare
$\Delta$ and $\pi$N pair contribution is model dependent.
Nevertheless, it is interesting to see how this is realized in a
typical case. Using the spectral function $\mathcal{A}$ shown in
fig.~\ref{phAB} the fraction of $\Delta$ resonances at freeze-out
is $n_{\Delta}/n_B = 0.035$ while that of $\Delta$'s and $\pi$N
pairs is $\tilde{n}_{\Delta}/n_B =0.054$. When the $\Delta$
resonance is treated as a stable particle ($\Gamma_\Delta = 0$) one
finds $n_{\Delta}/n_B = 0.037$ for the same values of $T$ and
$\mu_B$. In this case the contribution of the correlated $\pi$N
pairs vanishes.

We note that by arguments analogous to the ones presented here, one
finds that the invariant mass distribution of $\pi$N pairs \cite{eos}
emitted from a fireball should not be directly identified with a
thermally weighted $\Delta$ resonance. Again the correlated $\pi$N
pairs must be accounted for by using $\mathcal{B}$ rather than
$\mathcal{A}$.

\section{Summary and conclusions}

We have constructed the thermodynamic potential and the baryon density
of an interacting $\pi$N system taking the width of the $\Delta$
resonance consistently into account. At low densities, where medium
effects can be neglected, the weight function for the $\Delta$
resonance is given by its spectral function plus a correction
term. The latter accounts for the correlated $\pi$N pair component of
a physical $\Delta$ resonance. The full weight function
can, in agreement with the virial expansion, be expressed
in a model-independent way in terms of the energy derivative of the
$\pi$N scattering phase shift.

The frequently employed prescription, where the thermodynamics
involves a trivial spectral function for the nucleon
($\delta$-function) and the free spectral function of the $\Delta$
resonance, is incorrect. In such an approach, the contribution of
$\pi$N pairs is neglected. These pairs yield an important contribution
to the population of $\Delta$-like states and to the low-momentum pion
spectrum in relativistic heavy-ion collisions. When we include all
contributions, the thermal pions, those due to $\Delta$ decays, as
well as the contribution due to $\pi$N pairs, we find a good
description of the experimental $\pi^0$ spectrum of Au+Au at 1~GeV per
nucleon.

We thank  E.~Grosse, V.~Metag, H.~Oeschler, P.~Senger, C.~Sturm and
A.~Wagner for rewarding discussions.

\end{document}